\documentclass[showpacs,twocolumn,groupedaddress,pra,aps,
]{revtex4-1}

\usepackage{amsmath}
\usepackage{amssymb}
\usepackage{maybemath}
\usepackage{xcolor}
\usepackage{graphicx}
\usepackage{ulem}
\usepackage{bm}
\usepackage{comment}
\usepackage{braket}
\usepackage{mathrsfs}
\usepackage{tikz}

\usetikzlibrary{decorations.pathmorphing}
\usetikzlibrary{shapes}
\usetikzlibrary{shapes.geometric}
\usetikzlibrary{calc}

\tikzset{
    magnetic/.style={
        fill,
        shape border rotate=-90,
        isosceles triangle,
        isosceles triangle apex angle=60,
        node distance=1,
        minimum height=.1
    }
}

\tikzset{
    othermagnetic/.style={
        fill,
        shape border rotate=90,
        isosceles triangle,
        isosceles triangle apex angle=60,
        node distance=1,
        minimum height=.1
    }
}

\usepackage{dcolumn}
\newcolumntype{.}{D{.}{.}{-1}}

\usepackage[%
  colorlinks=true,
  urlcolor=blue,
  linkcolor=blue,
  citecolor=blue
]{hyperref}

\makeatletter

\newcommand{\Rmnum}[1]{\expandafter\@slowromancap\romannumeral #1@}
\makeatother

\begin{document}

\newcommand{\addrMPIK}{Max Planck Institute for Nuclear Physics, Saupfercheckweg 1, 69117 Heidelberg}

\title{Testing Standard Model extensions with few-electron ions}

\author{V. Debierre}
\email{vincent.debierre@mpi-hd.mpg.de}
\author{N.~S. Oreshkina}
\email{natalia.oreshkina@mpi-hd.mpg.de}
\author{I.~A. Valuev}
\author{Z. Harman}
\email{zoltan.harman@mpi-hd.mpg.de}
\author{C.~H. Keitel}
\affiliation{\addrMPIK}

\begin{abstract}
When collecting spectroscopic data on at least four isotopes, nonlinearities in the King plot are a possible sign of Physics beyond the Standard Model. In this work, an improved approach to the search for hypothetical new interactions with isotope shift spectroscopy of few-electron ions is presented. Very careful account is taken of the small nuclear corrections to the energy levels and the gyromagnetic factors, which cause deviations from King linearity within the Standard Model and are hence a possible source of confounds. In this new approach, the experimental King nonlinearity is not compared to the vanishing prediction of the Standard Model at the leading order, but to the calculated full Standard Model contribution to King nonlinearity. This makes searching for beyond-the-Standard-Model physics with King linearity analysis possible in a very-high-precision experimental regime, avoiding confounds. The bounds which can be set on beyond-the-Standard-Model parameters remain limited by the uncertainties on the small Standard Model nuclear corrections which cause King nonlinearity. Direct comparison between theory and experiment on a single pair of isotopes is advocated as a more suitable approach for few-electron ions.
\end{abstract}


\maketitle

\section{Introduction} \label{sec:Intro}

Recent years have seen the accelerating development of tests of Standard Model (SM) extensions at the precision frontier~\cite{CoulombHidden,LowEFrontier,BBPrec,PossibleForces,FifthForce,FlambaumIso,FifthForceG,NonLinXPYb,LinXPCa,bekker2019detection,Sailer2022}. In a context where the SM of fundamental physics is known to be incomplete, but where direct signals of new particles or interactions have so far not been reported, the precision frontier constitutes an interesting complement to the search for physics beyond the SM (also known as New Physics (NP)) at particle colliders, which probe the so-called energy and intensity frontiers. 
At the precision frontier,  rigorous searches for NP signatures can be carried out with relatively small-sized experiments. In particular, in atomic physics, very precise experiments are often supported by calculations. Such high-precision calculations are possible in particular for few electron-ions, and the development of bound-state quantum electrodynamics (QED) has allowed predictions of fundamental spectroscopic quantities with relative precisions as good as $3\times10^{-15}$ in the case of an optical transition \cite{Micke2020}, and $2.8\times10^{-11}$ in the case of the bound-electron $g$ (gyromagnetic) factor~\cite{Sturm13,Sturm14}.

Comparisons between experimental and theoretical results on fundamental atomic spectroscopic quantities, such as transition frequencies and $g$ factors, constitute stringent tests for QED in strong external fields (here, the electrostatic field of the nucleus). Such comparisons have also been used to determine fundamental constants, such as the electron rest mass~\cite{Sturm14}, or the proton charge radius~\cite{ProtonRadiusSolved}, with unprecedented precision. There  also have been proposals for an improved determination of the fine-structure constant $\alpha$ of electrodynamics~\cite{WDiffOld,GFactorAlpha,WDiffZ,HalilReduced, schneider2022direct}).

As has been shown in recent works, precision atomic spectroscopy can be used to set bounds on proposed SM extensions (conversely, it could be used to find signals of such beyond-SM phenomena). In particular, studies of the isotope shift of transition frequencies in singly-charged ions (and neutral atoms) have been shown~\cite{PossibleForces,FifthForce,FlambaumIso,NonLinXPYb,LinXPCa} to allow for the setting of stringent bounds on NP, with ``minimal theory inputs''. Isotope shift data can be checked for deviations from King planarity which, with some care, could be interpreted as signatures of NP. Care is important, because departures from King planarity are also caused by small nuclear corrections to spectroscopic quantities within the SM~\cite{FlambaumIso,VladimirKing,FifthForceG,NonLinXPYb}. A recent study on the transition frequencies of argon ions has shown that such effects had been underestimated in the past~\cite{VladimirKing,FifthForceG}. For a more systematically reliable interpretation of future experimental data on isotope shifts, it is important to estimate these small nuclear corrections accurately.

In the case of few-electron ions, calculations are generally more tractable than in the many-electron case, and hence can reach higher precisions. As such, these systems, and in particular their $g$ factor, could become prime testing grounds for proposed SM extensions, with joint efforts from the theory and the experiment sides. On the theory side, more accurate calculations of the energy levels, the $g$ factors, and their respective isotope shifts are made possible by the progress of bound-state QED. Both radiative QED contributions (two-loop radiative corrections)~\cite{VladZVP,CzarneckiLetter,CzarneckiLog,SikoraTwoLoop,TwoLoopMLoop} and nuclear contributions~\cite{FNSGHigher,RadiativeNuclSize,NuPolNum,NuDefJacek,NuDefNiklas,IgorSkyrme} (we will discuss them in more detail throughout this work) are under active consideration, and are being determined with increasing precision. On the experiment side, the ALPHATRAP group has reported the measurement of the $g$ factor of B-like Ar\textsuperscript{13+} as its first result~\cite{Sturm19}, with a precision of $1.4$ parts per billion, which represents an improvement on the previous value by seven orders of magnitude. It is projected~\cite{Sturm19} that the precision of future such measurements in Penning traps will outmatch the currently most precise reported results, which concern carbon and silicon (electron mass). Moreover, it is thought that not only $g$ factors, but differences of $g$ factors between two trapped ions, can be measured with a precision of one part in $10^{11}$ and better~\cite{Sailer2022}. This includes isotope shifts, which makes the present investigation very timely. High-precision measurements of the $g$ factor are also planned within the HITRAP project~\cite{Qui01,Herfurth_2015,Vogel2019}.

In this work, we study the setting of bounds on NP with few-electron ions, and show a simple approach based on direct comparison between theory and experiment to be best suited. This is in contrast to the case of many-electron ions, where a data-driven approach based on King representations and their generalisations is favoured~\cite{BerengutGenKing}. 
We present in detail an advanced King-type formalism to set bounds on NP through isotope shift spectroscopy of few-electron ions by taking careful account of the aforementioned small nuclear corrections and their contribution to King nonplanarity, and show that this complex formalism is suboptimal for few-electron ions, compared to a simpler approach. The rest of this manuscript is organized as follows. In Sec.~\ref{sec:Extensions}, we briefly present proposed extensions of the SM which can be addressed through atomic spectroscopy, and through isotope shifts in particular. Radiative and inter-electronic-interaction QED corrections to the NP contributions to the $g$ factor are discussed in Sec.~\ref{sec:QEDNP}. In Sec.~\ref{sec:Indirect}, the central part of this work, we study in detail the isotope shift of bound-electron energy levels and $g$ factors, and build a modified King formalism to search for NP with spectroscopic data on three isotope pairs. It is shown that this indirect approach, while well-suited for many-electron systems, is unnecessary and even suboptimal for few-electron systems.  In Sec.~\ref{sec:Direct}, we argue that direct comparisons between experimental and theoretical results on few-electron ions can be used to set more competitive bounds on NP with a single isotope pair. Sec.~\ref{sec:Ccl} is reserved for concluding remarks.

\section{Some extensions of the Standard Model} \label{sec:Extensions}

Several proposed extensions of the SM can be probed with atomic physics experiments. We briefly discuss these extensions in the present section.

\subsection{Higgs portal relaxion} \label{subsec:Higgs}

The long-standing electroweak hierarchy problem~\cite{EightPage}, which is connected to the lack of an explanation of the weakness of gravity with respect to the other forces, concerns the mass of the Higgs boson, which could be expected to be much larger than it is due to coupling of the Higgs boson with any beyond-SM phenomena. Indeed, radiative corrections to the Higgs mass vary as the square of the energy scale at which beyond-SM phenomena might arise. The Higgs portal provides a solution to this problem. It involves the mixing of a new massive scalar boson, the relaxion, with the Higgs boson. These massive scalar bosons would mediate a new fundamental force, resulting, as far as atomic physics is concerned, in an interaction between nucleons and electrons~\cite{FifthForce,FlambaumIso,ProbingIS}. The spin-independent potential exerted on electrons by this hypothetical force is of the Yukawa type~\cite{PossibleForces}:
\begin{equation} \label{eq:HPPotential}
  V_{\mathrm{HR}}\left(\mathbf{r}\right)=-\hbar c\,\alpha_{\mathrm{HR}}\,A\,\frac{\mathrm{e}^{-\frac{m_\phi c}{\hbar}\left|\mathbf{r}\right|}}{\left|\mathbf{r}\right|},
\end{equation}
where $m_\phi$ is the mass of the relaxion, $\alpha_{\mathrm{HR}}=y_ey_n/4\pi$ is the Higgs portal coupling constant, with $y_e$ and $y_n$ the coupling of the massive scalar boson to the electrons and the nucleons, respectively, $\hbar$ and $c$ are Planck's reduced constant and the vacuum velocity of light, and $A$ is the nuclear mass number of the considered ion.
\begin{figure}[b!]
\begin{center}
  \begin{tikzpicture}[very thick,scale=.625]
    \draw (-3.25,-.075) -- (-2.25,-.075);
    \draw (-3.25,.075) -- (-2.25,.075);

    \fill (-2.75,0) circle (.1);
    \fill (.75,0) circle (.1);

    \draw[dashed] (-2.75,.125)  -- (-2.75,1.25);


    \fill (-2.875,1.5) -- (-2.875,1.25) -- (-2.625,1.25) -- (-2.625,1.5) -- (-2.875,1.5);
    
    \draw (-3.25,2) node {($E$)};


    \draw (-1.25,-.075) -- (1.25,-.075);
    \draw (-1.25,.075) -- (1.25,.075);

    \fill (-.75,0) circle (.1);
    \fill (.75,0) circle (.1);

    \draw[decorate,decoration={snake,amplitude=1.5,segment length=6.25}] (.75,.125)  -- (.75,1.25);
    \draw[dashed] (-.75,.125)  -- (-.75,1.25);

    \draw[inner sep=2] (.75,1.375) node [magnetic]{};

    \fill (-.625,1.5) -- (-.625,1.25) -- (-.875,1.25) -- (-.875,1.5) -- (-.625,1.5);

    \draw (-1.25,2) node {($g$)};
    \draw (0,-.75) node {(2)};

  \end{tikzpicture}
\end{center}  
  \vspace{-15pt}
  \caption{Feynman diagram corresponding to the leading contribution to the hypothetical fifth force correction to the energy level ($E$) and the $g$ factor ($g$) of a bound electron. The double line represents the bound electron, the dashed line terminated by a square denotes the New Physics potential, and the wavy line terminated by a triangle denotes a photon from the external magnetic field. Diagram ($g$) has an equivalent diagram, as such, its contributions should be counted twice. \label{fig:FDiagrams5}}
\end{figure}
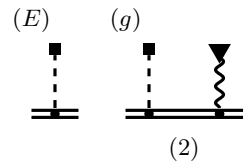

The first-order correction to the energy level of a bound electron in the quantum state $a$ due to this potential corresponds to the diagram in Fig.~\ref{fig:FDiagrams5}. The contribution is 
\begin{multline} \label{eq:ECorr}
E_{\mathrm{HR}\left(a\right)}=\langle a | V_{\mathrm{HR}} | a \rangle\  \\=-\alpha_{\mathrm{HR}}\,A\,\hbar c\int_0^{+\infty}\mathrm{d}r\,r\,\mathrm{e}^{-\frac{m_\phi c}{\hbar}r}\left[g_a^2\left(r\right)+f_a^2\left(r\right)\right].
\end{multline}
Here $g_a$ and $f_a$ are the radial wave functions (large and small component, respectively) of the bound electron in state $a$~\cite{Drake}. For the H-like ground state $a=1s$, we obtain 
\begin{multline} \label{eq:ENP}
  E_{\mathrm{HR}\left(1s\right)}  = -\alpha_{\mathrm{HR}}\,A\,m_e\,c^2\,\frac{\left(Z\alpha\right)}{\gamma}\left(1+\frac{m_\phi}{2Z\alpha m_e}\right)^{-2\gamma}
  \end{multline}
where $\gamma^2=1-\left(Z\alpha\right)^2$. 
{Here and below, we use the point-like-nucleus approximation for wavefunctions and energies to deliver analytical formulas.} The first-order correction to the $g$ factor of a bound electron in the quantum state $a$ due to this potential corresponds to the diagram in Fig.~\ref{fig:FDiagrams5}, together with the one in which the order of the two interactions is swapped. The contribution is
\begin{multline} \label{eq:GCorr}
  g_{\mathrm{HR}\left(a\right)}=-2\alpha_{\mathrm{HR}}\,A\,\frac{\hbar c}{\mu_0 B m_a}\int_0^{+\infty}\mathrm{d}r\,r\,\mathrm{e}^{-\frac{m_\phi c}{\hbar}r}\\
  \times\left[g_a\left(r\right)X_a\left(r\right)\vphantom{\left(\frac{1}{2m_e}\right)}+f_a\left(r\right)Y_a\left(r\right)\right],
\end{multline}
where $\mu_0=e\hbar/2m_e$ is the Bohr magneton, $m_a$ the magnetic projection quantum number of state $a$, $B$ the magnitude of the external, static, homogeneous magnetic field, and $X_a$ and $Y_a$ the corrections to the large and small components of the bound electron radial wave function, due to the interaction with the magnetic field, given in Ref.~\cite{ShabaevVirial}. With these wave function corrections and after performing the angular integration, 
Eq.~(\ref{eq:GCorr}) can be rewritten as 
\begin{widetext}
\begin{equation} \label{eq:GCorrEnd}
  g_{\mathrm{HR}\left(a\right)}=2\alpha_{\mathrm{HR}}\,A\,\lambdabar_e\frac{\kappa_a^2}{j_a\left(j_a+1\right)}\int_0^{+\infty}\mathrm{d}r\,r\,\mathrm{e}^{-\frac{m_\phi c}{\hbar}r}\left[\left(\frac{1}{2}-\kappa_a\right)g_a^2\left(r\right)+\left(\frac{1}{2}+\kappa_a\right)f_a^2\left(r\right)+2\frac{r}{\lambdabar_e}f_a\left(r\right)g_a\left(r\right)\vphantom{\left(\frac{1}{2}\pm\kappa_a\right)}\right],
\end{equation}
\end{widetext}
with $j_a$ and $\kappa_a$ the total and Dirac angular momentum numbers, respectively, and $\lambdabar_e=\hbar/2\pi m_e c$ the Compton wavelength of the electron.
For the H-like ground state $a=1s$, we obtain \cite{Sailer2022}
\begin{multline} \label{eq:GNP}
  g_{\mathrm{HR}\left(1s\right)}=-\frac{4}{3}\alpha_{\mathrm{HR}}\,A\,\frac{\left(Z\alpha\right)}{\gamma}\,\left(1+\frac{m_\phi}{2Z\alpha m_e}\right)^{-1-2\gamma}\\
  \times\left[1+\frac{m_\phi}{2Z\alpha m_e}\left(1+2\gamma\right)\right].
\end{multline}
It can be checked that the correction to the energy level and the $g$ factor obey the relation
\begin{equation} \label{eq:KLMId}
g_{\mathrm{HR}\left(1s\right)}=\frac{\kappa^2}{j\left(j+1\right)c^2}\frac{\partial E_{\mathrm{HR}\left(1s\right)}}{\partial m_e},
\end{equation}
which is derived in Ref.~\cite{ArbitraryPot} for arbitrary central potentials. The full exact results for the $a=2s$ ground state of Li-like ions, and for the $a=2p_{1/2}$ ground state of B-like ions, are given in the Supplemental Material to Ref.~\cite{FifthForceG}.

\subsection{Gauged $B-L$ symmetry} \label{subsec:BL}

The difference $B-L$ between the baryon and lepton numbers is conserved in the SM, and one proposed extension of the SM is built around the introduction of the $U\left(1\right)_{B-L}$ gauge symmetry. Unlike most group symmetries proposed to supplement the SM, this symmetry would be unbroken in the SM extension, giving rise to a new vector boson $Z'$, a massive hidden photon. Quantum anomalies in this model are canceled by the introduction of a right-handed neutrino for each lepton family, which introduces neutrino masses. In this SM extension, the boson $Z'$ couples electrons to nucleons with a Yukawa potential~\cite{PossibleForces,NuSignals}, meaning that the results of Ref.~\cite{FifthForceG}, recapitulated in Sec.~\ref{subsec:Higgs}, are directly applicable, by replacing the Higgs portal coupling $y_ey_n$ with the $B-L$ symmetry coupling $g_{B-L}^2$. In this case, the couplings of the boson to the neutron and electrons is identical, therefore, atomic physics is sufficient to resolve the individual couplings of massive hidden photons to SM particles, which was not the case for the relaxions studied in Sec.~\ref{subsec:Higgs}.

\subsection{Chameleon models} \label{subsec:Chameleon}

Chameleon particles have been proposed as dark energy candidates~\cite{ChameleonSurprise,ChameleonCosmology}. The most remarkable property of these scalar particles is that, due to their non-linear self-interaction, their mass is an increasing function of the energy density of the local environment. As such, in dense environments, these hypothetical particles would mediate a new force with a very limited range. Such models can be tested from atomic data. For practical purposes, the influence of this hypothetical particle on atomic spectra is described~\cite{BBPrec,TestChamGrav} by the potential
\begin{equation} \label{eq:ChamPotential}
  V_{\chi}\left(\mathbf{r}\right)=-\frac{1}{4\pi}\frac{\hbar c}{\left|\mathbf{r}\right|}\frac{m_e}{M_m}\left[\frac{M_A}{M_m}+\frac{1}{2}Z^2\alpha\frac{\hbar}{M_\gamma\,c\,\left|\mathbf{r}\right|}\right].
\end{equation}
Here $M_A$ is the mass of the nucleus of the considered ion, while $M_m$ and $M_\gamma$ are mass scales which are inversely proportional to the coupling strengths of the chameleon to matter and to electromagnetic energy density, respectively. Indeed, writing $\phi$ the chameleon field, $\rho$ the matter energy density, and $F^{\mu\nu}$ the Maxwell-Faray tensor, an effective field theory treatment yields the potential $V=\phi\,\rho/M_m$ for the chameleon-matter coupling potential, and $V=\phi\,F^{\mu\nu}F_{\mu\nu}/2M_m$ for the chameleon-photon coupling potential.

{In the following, we will use the case of Higgs relaxion portal for the explicit expressions, and refer to it as NP;  however our conclusions are also valid for the cases of gauged B-L symmetry and chameleon models, and can be analogously extended there. }
 
\section{QED corrections to the NP corrections} \label{sec:QEDNP}

NP contributions to spectroscopic quantities can be  expected to be very small, and QED corrections thereto should be even smaller (typically by a factor of $\alpha$ for radiative corrections at the one-loop level). But as was shown in Refs.~\cite{FNSVP, Glazov_2013}, the magnitude of such radiative QED corrections to a contribution from a potential can be comparable to, or even larger than the leading contribution from that potential, when the latter is generated by a highly localised potential. In our case, the leading contribution is the one-electron hypothetical NP correction, shown in Fig.~\ref{fig:FDiagrams5}. For the ground-state energy of H-like, Li-like and B-like ions, we have shown that radiative corrections to the NP contribution are much smaller than the leading NP contribution~\cite{PhotonBridge2021}. However, for heavy new bosons, photon-exchange corrections to the NP contribution to the energy levels of Li-like and B-like ions can be comparable or even much larger than the one-electron NP contribution~\cite{PhotonBridge2021}. We anticipate that the same could happen for the $g$ factor of Li-like and B-like ions.

\section{Indirect tests with the King approach} \label{sec:Indirect}

We now turn to the detailed study of isotope shifts of the energy levels and $g$ factors of highly charged ions. In Sec.~\ref{subsec:IsoShift}, we first present the standard, simple formulation of the isotope shift in spectroscopic data, and introduce the King representation of isotope shifts. In Sec.~\ref{subsec:Sublead}, we present in detail the various subleading nuclear corrections to the energy levels and to the $g$ factor, which contribute to the isotope shift, and complicate the analysis of data. Methods to distinguish SM and NP contributions to the isotope shift are discussed in Sec.~\ref{subsec:ModKing}. Projected bounds are derived in Sec.~\ref{subsec:Project}.

\subsection{Isotope shifts and the King representation} \label{subsec:IsoShift}

Isotope shift data is a promising avenue~\cite{FifthForce,FlambaumIso,PossibleForces,FifthForceG} to obtain strong bounds on NP parameters. This is easily understood when recalling that several proposed SM extensions would result in new forces between nucleons and electrons. Isotope shift data can thus carry information on potential neutron-electron interactions due to NP. Isotope shift data can conveniently be handled through the King representation~\cite{KingBook}. The bulk of the following explanation will be given for the explicit case of the $g$ factor but, as will be further clarified, the same construction can be elaborated for energy levels. Let us consider two levels $1$ and $2$ of a bound electron in an ion. For both levels, considering two isotopes $A$ and $A'$ of the same ion, we write~\cite{KingBook} the isotope shift in the $g$ factor as
\begin{equation} \label{eq:IsoShift}
g_i^{AA'}=g_i^A-g_i^{A'}\;,\quad i\in\left\{1,2\right\}.
\end{equation}
Within the SM, the leading-order (LO) contributions to the isotope shift are
\begin{equation} \label{eq:TradSM}
  g_{i\left(\mathrm{LO}\right)}^{AA'}=K_i\,\mu_{AA'}+F_i\,\delta R^2_{AA'}.
\end{equation}
The first summand on the r.h.s. of (\ref{eq:TradSM}) is the leading-order contribution to the mass shift, featuring the reduced mass $\mu_{AA'}=1/M_A-1/M_{A'}$. The second summand is the leading-order contribution to the field shift, featuring the difference $\delta R_{AA'}^2=R_A^2-R_{A'}^2$ (with $R\equiv\sqrt{\left\langle r^{2}\right\rangle}$) in the nuclear squared charge radii between the two isotopes. The reader should note that both summands are expressed as the product of an electronic and a nuclear factor. Indeed, the coefficients $K_i$ and $F_i$ only depend on the electronic level considered, while the quantities $\mu_{AA'}$ and $\delta R_{AA'}^2$ are purely nuclear properties.

As was done in Ref.~\cite{FifthForceG}, we introduce, for any pair of isotopes, the notation $G_i^{AA'}\equiv g_i^{AA'}/\mu_{AA'}$. For the pair of electronic levels $i$ and $j$, the King representation indicates the dependence of $G_j^{AA'}$ as a function of $G_i^{AA'}$. A King plot is a collection of points, with coordinates $\left(G_i^{AA'},G_j^{AA'}\right)$. To each pair $AA'$ of isotopes corresponds a point. From Eq.~(\ref{eq:TradSM}), it can easily be seen that
\begin{equation} \label{eq:Linear}
  G_{j\left(\mathrm{LO}\right)}^{AA'}=\frac{F_j}{F_i}G_{i\left(\mathrm{LO}\right)}^{AA'}+\left(K_j-\frac{F_j}{F_i}K_i\right).
\end{equation}
This relation between isotope shifts for two different levels is hence linear within the LO treatment, and the offset does not depend on the specific isotope pair considered. Hence, King plots will always be linear at the LO: points corresponding to any pair of isotopes will fall on the same line. Hence, deviations from King linearity in experimental data can either be explained by smaller, subleading nuclear contributions to the $g$ factor, or by possible NP contributions. We will explore this in detail in the following.
But first, we note that, in some cases, the relativistic correction to the field shift can be absorbed within this leading-order, linear framework. Indeed, in a relativistic treatment, the second summand on the r.h.s. of Eq.~(\ref{eq:TradSM}) can be rewritten as
\begin{equation} \label{eq:RelSM}
  g_{i\left(\mathrm{RLO}\right)}^{AA'}=K_i\,\mu_{AA'}+H_i\delta R_{AA'}^{2\gamma_i}.
\end{equation}
Here $\gamma_i=\sqrt{\kappa_i^2-\left(Z\alpha\right)^2}$, with $\kappa_i$ the relativistic angular quantum number, $\delta R_{AA'}^{2\gamma_i}=R_A^{2\gamma_i}-R_{A'}^{2\gamma_i}$,  and RLO stands for `relativistic leading order'. 
Obviously, the second summand on the r.h.s. of Eq.~(\ref{eq:RelSM}) is no longer the product of an electronic factor with a nuclear one. Nevertheless, if one considers only electronic levels that share the same $\left|\kappa_i\right|$, as we will do in this work, then $\gamma_i$ is in effect fixed, and, substituting $H_{i/j}$ for $F_{i/j}$, the linear relation (\ref{eq:Linear}) still holds 
between $G_{i\left(\mathrm{RLO}\right)}^{AA'}$ and $G_{j\left(\mathrm{RLO}\right)}^{AA'}$. 
As we will see in the upcoming Sec.~\ref{subsec:Sublead}, however, King linearity is broken by various subleading contributions to the isotope shift. Before we turn to these subleading corrections, we give, for reference, the expressions for the coefficients $K_i$ and $H_i$ for the levels $1s$, $2s$ and $2p_{1/2}$ which we consider in this work.

The mass shift coefficients are
\begin{widetext}
\begin{subequations} \label{eq:MassShift}
  \begin{align} 
    K_{1s}&=\left(Z\alpha\right)^2m_e\left[1-\frac{\left(Z\alpha\right)^2}{3}\left[1+\sqrt{1-\left(Z\alpha\right)^2}\right]^{-2}+\left(Z\alpha\right)^3P^{\left(1s\right)}\left(Z\alpha\right)\right], \label{eq:IGuessWeNeedIt}\\
    K_{2s}&=-\frac{2}{3}\frac{1}{m_e c^4}\left[2\kappa^2E_{2s}^2+\kappa\,m_e c^2\,E_{2s}-\left(m_e c^2\right)^2\right]+\frac{1}{8}m_e\left(Z\alpha\right)^5P^{\left(2s\right)}\left(Z\alpha\right),\\
    K_{2p_{1/2}}&=m_e\,F\left(Z\alpha\right).
\end{align}
\end{subequations}
These three results are derived, respectively, in Refs.~\cite{RecoilAllOrders,RecoilLithium,RecoilBoron}. Here $E_{2s}$ refers to the Dirac bound energy of the $2s$ level. The functions $P^{\left(1s\right)}$, $P^{\left(2s\right)}$ are tabulated in Refs.~\cite{RecoilAllOrders,RecoilLithium}, and the function $F$ is tabulated in Ref.~\cite{RecoilBoron} for $Z>20$.

The field shift coefficients are
\begin{subequations} \label{eq:FieldShift}
  \begin{align} 
    H_{1s}&=\frac{4}{3}\left(1+2\gamma\right)\frac{\left(Z\alpha\right)^2}{10}\left[1+\left(Z\alpha\right)^2f_{1s}\left(Z\alpha\right)\right] \left(2\sqrt{\frac{5}{3}}
    \frac{Z\alpha}{\lambdabar_e}\right)^{2 \gamma},\\
    H_{2s}&=\frac{4}{3}\left(1+2\gamma\right)\frac{\left(Z\alpha\right)^2}{20}\left[1+\left(Z\alpha\right)^2f_{2s}\left(Z\alpha\right)\right] \left(\sqrt{\frac{5}{3}}
    \frac{Z\alpha}{\lambdabar_e}\right)^{2 \gamma},\\
    H_{2p_{1/2}}&=\frac{1}{2}\left(1+2\gamma\right)\frac{\left(Z\alpha\right)^2}{40}\left[1+\left(Z\alpha\right)^2f_{2p_{1/2}}\left(Z\alpha\right)\right] \left(\sqrt{\frac{5}{3}}
    \frac{Z\alpha}{\lambdabar_e}\right)^{2 \gamma}
\end{align}
\end{subequations}
\end{widetext}
where $\lambdabar_e\equiv\hbar/m_ec$ is the Compton wavelength of the electron. The expressions for the functions $f_{1s}$, $f_{2s}$, $f_{2p_{1/2}}$ are given in Ref.~\cite{ShabaevSize}.

As was mentioned earlier, the foregoing considerations are also valid for the (dimensionless) energy levels $E_i/\left(m_e c^2\right)$ of bound electrons and their isotope shifts. In particular, for the ground-state energy of H-like ions, the coefficients are
\begin{equation} \label{eq:MassShiftE}
K_{1s}^{\left(E\right)}=\left(Z\alpha\right)^2m_e\left[\frac{1}{2}+\frac{\left(Z\alpha\right)^3}{\pi}P_0\left(Z\alpha\right)\right],
\end{equation}
with $P_0$ tabulated in Ref.~\cite{RecoilEnergy}, and
\begin{equation} \label{eq:FieldShiftE}
H_{1s}^{\left(E\right)}=\frac{\left(Z\alpha\right)^2}{10}\left[1+\left(Z\alpha\right)^2f_{1s}\left(Z\alpha\right)\right] \left(2\sqrt{\frac{5}{3}}
    \frac{Z\alpha}{\lambdabar_e}\right)^{2 \gamma}.
\end{equation}
Again, in this case, the relativistic correction to the field shift can be included directly, without breaking King linearity, as long as all the levels considered share the same~$\gamma_i$. Note that (also see Ref.~\cite{ArbitraryPot})
\begin{equation} \label{eq:HRelation}
H_{1s}=\frac{4}{3}\left(1+2\gamma\right)H_{1s}^{\left(E\right)},
\end{equation}
which inspired the introduction in Ref.~\cite{HalilReduced} of the reduced $g$ factor 
\begin{equation} \label{eq:ReducedG}
g_{1s}-x\frac{E_{1s}}{m_e c^2},\hspace{50pt}x\equiv\frac{4}{3}\left(1+2\gamma\right)
\end{equation}
of H-like ions, in which the leading finite nuclear size correction cancels to a great degree. As we will see, for several subleading nuclear corrections, the same relation obtains at least approximately, with the same proportionality factor $x$. This motivates the introduction of a King representation where the isotope shifts of the energy level and $g$ factor of the ground state of H-like ions are expressed as a function of each other.

\subsection{Subleading nuclear corrections and Standard-Model King nonplanarities} \label{subsec:Sublead}

Let us introduce an extra term in the isotope shift. As was done in the previous Sec.~\ref{subsec:IsoShift}, we write the general abstract expressions for the $g$ factor, but an identical construction can be, and indeed is, made for the energy levels. The isotope shift now reads
\begin{equation} \label{eq:FullSM}
  g_{i\left(\mathrm{SM}\right)}^{AA'}=K_i\,\mu_{AA'}+H_i\,\delta R_{AA'}^{2\gamma_i}+s_{i\,AA'}.
\end{equation}
Here SM stands for `Standard Model', as $s_{i\,AA'}$ encompasses all the SM other contributions to the isotope shift, besides the leading-order mass shift, and the relativistic leading-order field shift. The third summand $s_{i\,AA'}$ on the r.h.s. is thus generated by subleading nuclear corrections to the $g$ factor. We will examine these various corrections in detail in the following. For now, we can handle this term in a more abstract way, and start by noting that it is not the product of an electronic and a nuclear factor. Hence, the relation between isotope shifts for two electronic levels is
\begin{multline} \label{eq:nonlinearSM}
  G_{j\left(\mathrm{carrSM}\right)}^{AA'}=\frac{H_j}{H_i}G_{i\left(\mathrm{SM}\right)}^{AA'}+\left(K_j-\frac{H_j}{H_i}K_i\right)\\
  +\frac{1}{\mu_{AA'}}\left(s_{j\,AA'}-\frac{H_j}{H_i}s_{i\,AA'}\right).
\end{multline}
There is now an extra offset, in the relation between the isotope shifts of two electronic levels. This offset depends on the specific isotope pair considered, which means that the King plot will no longer be linear. Typically, though, the subleading terms $s_{i\,AA'}$ are much smaller than the other two summands on the r.h.s. of (\ref{eq:FullSM}), so that the deviations from King linearity due to these terms are only detected on isotope shift data if that data is sufficiently precise.

We now turn to the specific determination of the term $s_{i\,AA'}$. Our analysis includes the seven largest subleading nuclear corrections to the energy level and the $g$ factor: {1.~}the higher-order nuclear recoil and {2.~}nuclear size corrections, {3.~}the mixed finite-size recoil correction, {4.~}the radiative recoil and {5.~}radiative nuclear size corrections, as well as {6.~}the nuclear deformation and {7.~}nuclear polarization corrections. For reasons that will become clear, we will devote specific attention to the theoretical uncertainties on these various contributions. We will focus, from this point onwards, on the $g$ factor and the (dimensionless) ground-state energy level of H-like ions. A representation of the contributions from all seven considered nuclear corrections, to a specific isotope shift in H-like calcium, is given in Fig.~\ref{fig:RadarRed}. In the case of ions with more than one electron, nuclear recoil and finite nuclear size corrections to the inter-electronic interaction would also contribute to King nonlinearities, but we do not consider this case in further detail here.

\subsubsection{Higher-order nuclear recoil correction} \label{subsubsec:HORecoil}

The leading-order nuclear recoil correction expressed by Eqs.~(\ref{eq:MassShift}) and (\ref{eq:MassShiftE}) gives a contribution to the bound-electron $g$ factor proportional to $m_e/M_A$, where $m_e$ is the electron mass, and $M_A$ the nuclear mass (for a specific isotope). A result valid to all orders in the electron-nucleus mass ratio can be obtained, which does not take account of radiative or electron-electron interaction contributions. The result is given by Eqs.~(29)-(30) of Ref.~\cite{EidesGrotch} (also see Ref.~\cite{HOMass}); one should, to obtain the higher-order nuclear recoil correction, substract from it the leading-order nuclear recoil term $\left(m_e/M_A\right)\left(Z\alpha\right)^2/n^2$. The result of Ref.~\cite{EidesGrotch} is valid, as we said, at all orders in $m_e/M_A$, but is given as an expansion in $\left(Z\alpha\right)$. Inspection indicates that the next, {unacccounted-for} terms should feature an extra $\left(Z\alpha\right)^2$, which is how we estimate the uncertainty on this correction.

For the energy level, the higher-order nuclear recoil correction is obtained from Eqs.~(48) and (49) of Ref.~\cite{VladVladTable}. The unaccounted term should be of order $\left(Z\alpha\right)^6\left(m_e/M_A\right)^2$, which yields the estimate of the uncertainty on this correction.

\subsubsection{Higher-order finite nuclear size correction} \label{subsubsec:HOSize}

The leading-order finite-nuclear size correction expressed by Eqs.~(\ref{eq:FieldShift}) and (\ref{eq:FieldShiftE}) should be supplemented by smaller, subleading contributions. Calculation of these contributions can be achieved by numerically calculating the full finite-nuclear size correction to the $g$ factor and energy level, e.g. following the method of Ref.~\cite{IgorSkyrme}, and subtracting the already-included leading-order contribution. The uncertainty on these contributions is dominated by that on the nuclear radius.

\subsubsection{Finite-size recoil correction} \label{subsubsec:FNSRecoil}

As mentioned in Ref.~\cite{VladimirKing}, the nonrelativistic approximation of the finite-size recoil correction to the energy levels can be obtained through the substitution $H_i^{\left(E\right)}\rightarrow H_i^{\left(E\right)}\left(1-3\frac{m_e}{M_A}\right)$. Hence, following Ref.~\cite{ArbitraryPot}, we have, in the nonrelativistic approximation, $H_i\rightarrow H_i\left(1-4\frac{m_e}{M_A}\right)$ for the finite-size recoil correction to the $g$ factors of the levels under study here. The uncertainties in this estimate are given by multiplying this nonrelativistic estimate by $\left(Z\alpha\right)^2$, however, finite-size recoil corrections to the H-like energy level were computed with three significant digits in Ref.~\cite{RecoilEnergy}, so that we can retain this latter level of uncertainty as very realistic.

\subsubsection{Radiative corrections to the nuclear recoil correction} \label{subsubsec:RadRec}

Radiative (QED) corrections to the nuclear recoil contribution are mixed QED+nuclear corrections. For the $g$ factor of H-like ions, they are given by~\cite{ShabaevReview}
\begin{equation} \label{eq:RadiativeRecoilG}
\Delta g_{1s}^{\mathrm{rad.~rec.}}=\left(\frac{\alpha}{\pi}\right)\left(Z\alpha\right)^2\left[-\frac{1}{3}\frac{m_e}{M_A}+\frac{3-2Z}{6}\left(\frac{m_e}{M_A}\right)^2\right].
\end{equation}
Similarly, the radiative recoil correction to the energy level is given by ~\cite{VladVladTable}
\begin{multline} \label{eq:RadiativeRecoilE}
\frac{\Delta E_{1s}^{\mathrm{rad.~rec.}}}{m_e c^2}=\left(\frac{\alpha}{\pi}\right)\frac{\left(Z\alpha\right)^5}{\pi}\left[\frac{m_e}{M_A}-3\left(\frac{m_e}{M_A}\right)^2\right]\\
\times\left[6\zeta\left(3\right)-2\pi^2\log\left(2\right)+\frac{35}{26}\pi^2\right.\\
\left.-\frac{448}{27}-\frac{2}{3}\pi\left(Z\alpha\right)\log^2\left(\left(Z\alpha\right)^2\right)\right].
\end{multline}
Clearly, when considering the isotope shift, the first summand on the r.h.s. of Eqs.~(\ref{eq:RadiativeRecoilG}) and (\ref{eq:RadiativeRecoilE}), linear in $m_e/M_A$, will bring a small correction to the coefficient $K_i$, and hence no King nonlinearity. The terms quadratic in $\left(m_e/M_A\right)^2$ are extremely small, but taken into account in our treatment. Their uncertainties are estimated by multiplying the quadratic mass ratio term by $\left(Z\alpha\right)^2$ in the case of the $g$ factor (with Eq.~(\ref{eq:MassShift}) in mind), and by $\left(Z\alpha\right)$ in the case of the energy level (with Eq.~(\ref{eq:MassShiftE}) in mind).

\subsubsection{Radiative corrections to the finite size} \label{subsubsec:RadFNS}

The most detailed study of radiative nuclear size corrections to the $g$ factor and energy of H-like ions are found in Refs.~\cite{RadiativeNuclSize,RadiativeNuclSizeLamb}, respectively. The correction to the $g$ factor is cast as
\begin{equation} \label{eq:RadiativeSizeG}
\Delta g_{1s}^{\mathrm{rad.~FNS}}=\frac{8}{3}\left(\frac{\alpha}{\pi}\right)\left(Z\alpha\right)^4G_{\mathrm{NS}}\times G_{\mathrm{NSQED}}\left(\frac{R}{\lambdabar_e}\right)^{2\gamma}.
\end{equation}
The correction to the energy level  can be recast as
\begin{equation} \label{eq:RadiativeSizeE}
\frac{\Delta E_{1s}^{\mathrm{rad.~FNS}}}{m_e c^2}=\frac{2}{3}\left(\frac{\alpha}{\pi}\right)\left(Z\alpha\right)^4D_{\mathrm{NS}}\times D_{\mathrm{NSQED}}\left(\frac{R}{\lambdabar_e}\right)^{2\gamma}.
\end{equation}
The coefficients where $G_{\mathrm{NS}}$ and $G_{\mathrm{NSQED}}$, as well as $D_{\mathrm{NS}}$ and $D_{\mathrm{NSQED}}$ are functions of $Z$ and of the nuclear radius $R$, and are numerically calculated coefficients, with $G_{\mathrm{NS}}\simeq1$ (within $20\%$) and  $-2<G_{\mathrm{NSQED}}<-0.4$ for $Z\leq92$. The weighted difference only cancels the radiative nuclear size correction quite weakly (less than one-digit cancellation). When considering the isotope shift, it should be kept in mind that all of $R$, $G_{\mathrm{NS}}$ and $G_{\mathrm{NSQED}}$ are isotope-specific. In principle, the uncertainties on these radiative nuclear size corrections is only limited by the knowledge of the nuclear radii. This indicates that the relative uncertainties on the radiative nuclear size corrections can be as small as roughly one part pert thousand.

Finally, the two-loop radiative corrections to the finite nuclear size, and to the nuclear recoil contributions, are extremely small, of course, and can be ignored altogether.

\subsubsection{Nuclear deformation correction} \label{subsubsec:NuDef}

For nonspherical nuclei, which represent most of the nuclei, there exists a nuclear deformation correction (also called nuclear shape correction) to the $g$ factor. Typically, it is considered that nuclei possess a quadrupole and a hexadecapole deformation, although they may also exhibit an octupole deformation~\cite{NuclDefValues}. A {nonperturbative} calculation of the nuclear deformation correction to the $g$ factor has also been carried out, and presented in Ref.~\cite{NuDefNiklas}. In this approach, the wave functions for the bound electrons are obtained by numerically solving the Dirac equation for the potential generated by a deformed nucleus. It is found that, taking into account nuclear deformation at all orders modifies the results obtained through the perturbative method of Ref.~\cite{NuDefJacek}, to a relatively large extent. Nevertheless, the nuclear shape corrections to the $g$ factor and energy obey $\Delta g_{1s}^{\mathrm{NS}}=x\Delta E_{1s}^{\mathrm{NS}}/\left(m_e c^2\right)$ to a very good level of approximation, leading to a cancellation of about two digits in the reduced $g$ factor, and in the uncertainties, which is very favourable for our purposes.

\subsubsection{Nuclear polarization correction} \label{subsubsec:NuPol}

The bound electron and the nucleus can exchange photons, which excite virtual nuclear transitions from and to the nuclear ground state~\cite{NuPolAna,NuPolNum}, giving rise to the nuclear polarization correction. The detailed expressions for the corresponding correction to the bound-electron $g$ factor can be found in Refs.~\cite{HalilReduced,NuPolAna}. It has been shown through explicit numerical calculations that $\Delta g_{1s}^{\mathrm{NPOL}}\simeq x\Delta E_{1s}^{\mathrm{NPOL}}/\left(m_e c^2\right)$ to a good level of approximation~\cite{HalilReduced}, leading to a cancellation of at least one digit in the reduced $g$ factor difference.

\subsection{A modified King representation} \label{subsec:ModKing}

As we just discussed in detail, small nuclear corrections to the $g$ factor and the energy levels cause violations of King planarity within the SM. This limits the range of the test of NP by inspection of King plot data used in Ref.~\cite{FifthForce}. In Ref.~\cite{FifthForceG}, we used the method of Ref.~\cite{FifthForce} to study tests of NP with the bound-electron $g$ factor, while also making sure that the sensitivity to NP at the projected level of precision for potential Penning trap experiments would not be confounded by these subleading nuclear corrections. In the present work, we propose an extension of this method, which circumvents its fundamental limitation: namely, we propose to test the King nonlinearity of experimental data against its predicted theoretical value within the SM, instead of testing it against the zero nonlinearity predicted by the leading-order SM contributions to the isotope shift (see Sec.~\ref{subsec:IsoShift}). In all approaches, we need to consider the hypothetical NP contribution to King nonlinearity.

In the presence of NP, the isotope shift reads
\begin{equation} \label{eq:FullNP}
  g_i^{AA'}=K_i\,\mu_{AA'}+H_i\,\delta R_{AA'}^{2\gamma_i}+s_{i\,AA'}+n_{i\,AA'},
\end{equation}
where the fourth summand on the r.h.s. corresponds to the hypothetical NP contribution to the $g$ factor. The same general expression holds for the (dimensionless) energy levels. The minimal number of isotopes to consider to study King plots is four. As such, it is interesting to retain the general formalism developed in Ref.~\cite{FifthForce}, even though it is not directly applicable to data sets with more isotopes. We note that, in any case, little high-precision data on the isotope shift of $g$ factors is available, with the notable exception of Ref.~\cite{CalciumIS}, which motivates the use of this method, adapted to the case with the smallest number of isotope shift data points, that is sufficient to assess King planarity. We recall the definition $G_i^{AA'}\equiv g_i^{AA'}/\mu_{AA'}$. We also introduce $\delta^{AA'}\equiv\delta R_{AA'}^{2\gamma}/\mu_{AA'}$ \footnote{We recall that in this work, all considered electronic levels have the same $\gamma$}, $S_{i\,AA'}\equiv s_{i\,AA'}/\mu_{AA'}$ and $N_{i\,AA'}\equiv n_{i\,AA'}/\mu_{AA'}$. We now consider four isotopes $A$, $A'_1$, $A'_2$ and $A'_3$ and define the vectors 
\begin{widetext}
\begin{multline} \label{eq:Vectors}
  \begin{array}{r@{\hskip 0.04125in}c@{\hskip 0.04125in}l@{\hskip 0.625in}r@{\hskip 0.04125in}c@{\hskip 0.04125in}l@{\hskip 0.625in}r@{\hskip 0.04125in}c@{\hskip 0.04125in}l}
    \mathbf{G}_i&\equiv&\left(G_i^{AA'_1},G_i^{AA'_2},G_i^{AA'_3}\right),&
    \bm{\delta}&\equiv&\left(\delta^{AA'_1},\delta^{AA'_2},\delta^{AA'_3}\right),\\
    \mathbf{S}_i&\equiv&\left(S_{i\,AA'_1},S_{i\,AA'_2},S_{i\,AA'_3}\right),&
    \mathbf{N}_i&\equiv&\left(N_{i\,AA'_1},N_{i\,AA'_2},N_{i\,AA'_3}\right),&
    \mathbf{A}_{\mathrm{tt}}&\equiv&\left(1,1,1\right).
\end{array}
\end{multline}
Note, for future purposes, that for the Higgs portal, $\mathbf{N}_i=\alpha_{\mathrm{HP}}X_i\mathbf{h}$, with
\begin{equation} \label{eq:HVect}
\mathbf{h}\equiv\left(\frac{A-A'_1}{\mu_{AA'_1}},\frac{A-A'_2}{\mu_{AA'_2}},\frac{A-A'_3}{\mu_{AA'_3}}\right).
\end{equation}
In the rest of this work, we thus give expressions for the Higgs portal, to be explicit, but these expressions are straightforwardly applied to the case of the unbroken $B-L$ symmetry when substituting the coupling constant $4\pi g_{B-L}^2$ for $\alpha_{\mathrm{HP}}$, and, as can be seen from Eq.~(\ref{eq:ChamPotential}), by substituting $m_e m_{p/n}/4\pi M_m^2$. King nonplanarity (or nonlinearity) is measured~\cite{FifthForce} by the parameter
\begin{equation} \label{eq:Non}
  \mathcal{N}\equiv\frac{1}{2}\left(\mathbf{G}_1\times\mathbf{G}_2\right)\cdot\mathbf{A}_{\mathrm{tt}},
\end{equation}
which can be calculated to be equal to
  \begin{equation} \label{eq:Argh}
    \mathcal{N}=\frac{1}{2}\left(H_1\bm{\delta}\times\mathbf{S}_2-H_2\bm{\delta}\times\mathbf{S}_1+H_1\bm{\delta}\times\mathbf{N}_2-H_2\bm{\delta}\times\mathbf{N}_1+\mathbf{S}_1\times\mathbf{S}_2+\mathbf{S}_1\times\mathbf{N}_2-\mathbf{S}_2\times\mathbf{N}_1+\mathbf{N}_1\times\mathbf{N}_2\right)\cdot\mathbf{A}_{\mathrm{tt}}.
  \end{equation}
\end{widetext}
Note that if $n_{i\,AA'}$ is expressed as the product of $A-A'$ with a purely electronic factor, as is the case for the Higgs portal and the gauged $B-L$ symmetry, then $\mathbf{N}_1\times\mathbf{N}_2=\mathbf{0}$. If we neglect the subleading nuclear contributions to the isotope shift ($\mathbf{S}_i=\mathbf{0}$), and consider that there are no NP contributions ($\mathbf{N}_i=\mathbf{0}$), then it is seen from Eq.~(\ref{eq:Argh}) that King nonplanarity will vanish. Hence, nonplanar King data might be interpreted as a sign of NP, provided that this nonplanarity is not expected to arise from the subleading nuclear contributions. Experimentally, the presence or absence of nonplanarity can only be assessed at a certain level of accuracy, since, due to experimental uncertainties, we could not expect King data to verify $\mathcal{N}=0$ exactly, even in a \textit{Gedankenexperiment} in which $\mathbf{S}_i=\mathbf{N}_i=\mathbf{0}$. We propose here to introduce a modified King nonplanarity, that accounts for (and hence, in a way, sets aside) the deviation from the nonplanarity expected within the SM. This nonplanarity is obtained by a combination of experimental data and theoretical calculations of subleading nuclear corrections, and reads:
\begin{align} \label{eq:ModNon}
  \mathcal{N}'\equiv&\frac{1}{2}\left(\mathbf{G}_1\times\mathbf{G}_2\right)\cdot\mathbf{A}_{\mathrm{tt}}\nonumber\\
  -&\frac{1}{2}\left(H_1\bm{\delta}\times\mathbf{S}_2-H_2\bm{\delta}\times\mathbf{S}_1+\mathbf{S}_1\times\mathbf{S}_2\right)\cdot\mathbf{A}_{\mathrm{tt}}.
\end{align}
The first summand can be obtained experimentally, by collecting isotope shift data. The second summand can be calculated theoretically with the methods summarised in Sec.~\ref{subsec:Sublead}. The new nonplanarity test is performed as follows. The modified nonplanarity parameter (\ref{eq:ModNon}) is first compared to its first-order propagated error $\sigma_{\mathcal{N}'}$. If $\left|\mathcal{N}'\right|<\sigma_{\mathcal{N}'}$, the data is considered planar. Then, the NP parameter which is to be constrained, is considered to be bound by its own first-order propagated error, which can be computed from Eq.~(\ref{eq:Argh}). Note that the error $\sigma_{\mathcal{N}'}$ features not only experimental (as in Refs.~\cite{FifthForce,ProbingIS,FifthForceG}) but also theoretical contributions. Indeed, the subleading nuclear contributions $\mathbf{S}_i$ to the isotope shift have uncertainties. The explicit expression of the error is:
\begin{align} \label{eq:PropError}
  \sigma_{\mathcal{N}'}&=\sqrt{\sum_{i=1}^2\sum_{j=1}^3\left(\frac{\partial\mathcal{N}'}{\partial G_{i\left(j\right)}}\right)^2\left(\Delta G_{i\left(j\right)}\right)^2}\nonumber\\
  &+\sqrt{\sum_{i=1}^2\sum_{j=1}^3\left(\frac{\partial\mathcal{N}'}{\partial S_{i\left(j\right)}}\right)^2\left(\Delta S_{i\left(j\right)}\right)^2}
\end{align}
where $G_{i\left(j\right)}$ (resp. $S_{i\left(j\right)}$) is the $j$ component of $\mathbf{G}_i$ (resp. of $\mathbf{S}_i$), which is an experimental (resp. theoretical) quantity.

This modified King planarity test can detect nonplanarities caused by NP contributions to the $g$ factor, even when these nonplanarities are much smaller than those caused by the subleading nuclear corrections to the $g$ factor. This can be seen as follows. Let us write $\mathbf{E}_i$ the (signed) difference between the measured isotope shifts $\mathbf{G}_i$ and the corresponding SM predictions, obtained through Eq.~(\ref{eq:FullSM}). In the very high precision regime of experiments, where $\mathbf{E}_i$ is expected to be smaller than the subleading nuclear corrections $\mathbf{S}_i$, we can approximate $\sigma_{\mathcal{N}'}$ by its second summand in Eq.~(\ref{eq:PropError}), that is, by
\begin{widetext}
\begin{equation} \label{eq:LeviCivitaError}
  \sigma_{\mathcal{N}'}\simeq\frac{1}{2}\left[\sum_{i=1}^2\sum_{j=1}^3\left(\sum_{k=1}^3\sum_{l=1}^3\epsilon_{jkl}\left(H_{\bar{\imath}}\delta_{\left(l\right)}+S_{\bar{\imath}\left(l\right)}\right)\right)^2\left(\Delta S_{i\left(j\right)}\right)^2\right]^{\frac{1}{2}}
\end{equation}
where $\delta_{\left(i\right)}$ is the $i$ component of $\bm{\delta}$; and $\bar{\imath}=3-i$. This first-order propagated error is to be compared to 
\begin{equation} \label{eq:LeviCivitaPlain}
  \mathcal{N}'=-\frac{1}{2}\sum_{i=1}^2\sum_{j=1}^3\sum_{k=1}^3\sum_{l=1}^3\epsilon\left(i\right)\epsilon_{jkl}\left(H_{\bar{\imath}}\delta_{\left(k\right)}E_{i\left(l\right)}+\frac{1}{2}E_{i\left(k\right)}E_{\bar{\imath}\left(l\right)}+S_{\bar{\imath}\left(k\right)}E_{i\left(l\right)}\right)
\end{equation}
\end{widetext}
where $E_{i\left(j\right)}$ is the $j$ component of $\mathbf{E}_i$ and $\epsilon\left(i\right)=\left(-1\right)^{1+i}$. If there are no NP contributions, the deviations $\mathbf{E}_i$ between the experimental results and the SM predictions can be expected to become smaller with improving experimental precision. In this case, $\left|\mathcal{N}'\right|$ decreases too, and we can expect to reliably have $\left|\mathcal{N}'\right|<\sigma_{\mathcal{N}'}$, which allows one to set bounds on the NP parameters. If the theory of the nuclear corrections to the $g$ factor improves, then the uncertainties $S_{i\left(j\right)}$ will decrease, causing a similar decrease in $\sigma_{\mathcal{N}'}$. Nevertheless, if there are no NP contributions to the $g$ factor, $\mathbf{E}_i$ will also decrease with improving theoretical precision, as the theory will better match the experiments. This means that, within this programme, there is in principle no limit of applicability of the King planarity test, contrasting with the test used in Refs.~\cite{FifthForce,FifthForceG}.

\subsection{Projected bounds} \label{subsec:Project}

To derive bounds, we recast the (standard) King nonplanarity as~\cite{FifthForceG}
\begin{widetext}
\begin{equation} \label{eq:ArghNArrows}
  \mathcal{N}=-\frac{\mathbf{A}_{\mathrm{tt}}}{2}\cdot\left[\mathbf{G}_1\times\left(\frac{H_2}{H_1}\mathbf{S}_1-\mathbf{S}_2\right)-\left(1\leftrightarrow2\right)\right]+\frac{\alpha_{\mathrm{HP}}}{2}\left(\mathbf{A}_{\mathrm{tt}}\times\mathbf{h}\right)\cdot\left[\left(\frac{H_2}{H_1}X_1-X_2\right)\mathbf{G}_1-\left(1\leftrightarrow2\right)\right].
\end{equation}
In the following, quantities with index $i=1$ will refer to the H-like $g$ factor, and those with index $i=2$ will refer to the dimensionless H-like ground-state energy. Our King representation will be explicitly constructed, to compare the isotope shifts of these two quantities. This expression can be inverted to obtain
\begin{equation} \label{eq:SolveFor}
\alpha_{\mathrm{HP}}=2\frac{\left[\mathcal{N}+\frac{\mathbf{A}_{\mathrm{tt}}}{2}\cdot\left[\mathbf{G}_1\times\left(\frac{H_2}{H_1}\mathbf{S}_1-\mathbf{S}_2\right)-\left(1\leftrightarrow2\right)\right]\right]}{\left(\mathbf{A}_{\mathrm{tt}}\times\mathbf{h}\right)\cdot\left[\left(\frac{H_2}{H_1}X_1-X_2\right)\mathbf{G}_1-\left(1\leftrightarrow2\right)\right]}.
\end{equation}
\end{widetext}
Now, note that while studying the various subleading nuclear corrections to the $g$ factor and the energy levels, we have mentioned that, for several of them, the equality $\Delta g_{1s}\simeq x\Delta E_{1s}/\left(m_e c^2\right)\simeq0$ holds in good approximation. Hence, we can write
\begin{equation}
\mathbf{S}_i=\mathbf{T}_i+\mathbf{U}_i,\hspace{50pt}\mathbf{T}_1=x\,\mathbf{T}_2
\end{equation}
with $\mathbf{T}_i$ coming from the subleading nuclear corrections for which $\Delta g_{1s}\simeq x\Delta E_{1s}/\left(m_e c^2\right)\simeq0$, and $\mathbf{U}_i$ from the other corrections. Fortunately, $H_2/H_1 X_1$ and $X_2$, which appears in the denominator on the r.h.s. of Eq.~(\ref{eq:SolveFor}), does not cancel to any appreciable degree except in the very large boson mass limit (as can be checked by Eqs.~(\ref{eq:ENP}), (\ref{eq:GNP}), (\ref{eq:FieldShift}) and (\ref{eq:FieldShiftE})). The bounds on $\alpha_{\mathrm{HP}}$ are given by the first-order propagated error
\begin{align} \label{eq:PropErrorAlpha}
  \sigma_{\alpha_{\mathrm{HP}}}&=\sqrt{\sum_{i,j}\left(\frac{\partial\alpha_{\mathrm{HP}}}{\partial G_{i\left(j\right)}}\right)^2\left(\Delta G_{i\left(j\right)}\right)^2}\nonumber\\
  &+\sqrt{\sum_{i,j}\left(\frac{\partial\alpha_{\mathrm{HP}}}{\partial S_{i\left(j\right)}}\right)^2\left(\Delta S_{i\left(j\right)}\right)^2}
\end{align}
where the derivatives can be obtained from Eq.~(\ref{eq:SolveFor}). We repeat that here, $\mathbf{G}_1$ and $\mathbf{G}_2$ are to be understood as purely experimental quantities, and so too is $\mathcal{N}$, which is given in terms of these two vectors by Eq.~(\ref{eq:Non}). The $\Delta G_{i\left(j\right)}$ are hence experimental uncertainties. On the other hand, $\mathbf{S}_1$ and $\mathbf{S}_2$ are to be calculated, and the $\Delta S_{i\left(j\right)}$ are theoretical uncertainties. They are the uncertainties of the subleading nuclear corrections.

It is important to remember that $H_1=x\,H_2$. As can then be seen by inspecting the r.h.s. of Eq.~(\ref{eq:SolveFor}), this means that, for all occurrences of $\mathbf{S}_i$ in Eqs.~(\ref{eq:SolveFor}) and (\ref{eq:PropErrorAlpha}), the corresponding contribution will be dominated by $\mathbf{U}_i$, the contribution from the subleading corrections that do not cancel appreciably through the reduced $g$ factor. On Fig.~\ref{fig:RadarBlue}, we show the contributions to the predicted SM King nonplanarity in the vanishing boson mass limit from all seven considered subleading nuclear corrections. We also show the contributions, from these same corrections, to the bound on the NP coupling constant. It is seen that the largest contributors to the theoretical uncertainty budget are the higher-order nuclear recoil and nuclear size contributions and, to a lesser extent, the nuclear polarization correction. The latter is currently being investigated with unprecedentedly accurate nuclear  models~\cite{IgorSkyrme,Valuev2022}, and it is anticipated that the calculation uncertainties will be significantly reduced in the near future. It is more difficult to substantially improve the calculation of the higher-order recoil correction.
\begin{figure*}
\begin{center}
  \begin{tikzpicture}[thin,>=stealth,scale=.625]
    
    \foreach \x in {1,2,3,4,5,6,7}
    \draw (0,0) -- ({5*cos((\x*360/7)+90)},{5*sin((\x*360/7+90))});
    
    \foreach \x in {1,2,3,4,5}
    \draw ({\x*cos((1*360/7)+90)},{\x*sin((1*360/7+90))}) -- ({\x*cos((2*360/7)+90)},{\x*sin((2*360/7+90))}) -- ({\x*cos((3*360/7)+90)},{\x*sin((3*360/7+90))}) -- ({\x*cos((4*360/7)+90)},{\x*sin((4*360/7+90))}) -- ({\x*cos((5*360/7)+90)},{\x*sin((5*360/7+90))}) -- ({\x*cos((6*360/7)+90)},{\x*sin((6*360/7+90))}) -- ({\x*cos((7*360/7)+90)},{\x*sin((7*360/7+90))}) -- ({\x*cos((1*360/7)+90)},{\x*sin((1*360/7+90))});
    
    \draw (0,{sin((4*360/7)+90))-.25}) node {$18$};
    \draw (0,{2*sin((4*360/7)+90))-.25}) node {$16$};
    \draw (0,{3*sin((4*360/7)+90))-.25}) node {$14$};
    \draw (0,{4*sin((4*360/7)+90))-.25}) node {$12$};
    \draw (0,{5*sin((4*360/7)+90))-.25}) node {$10$};
    
    \draw ({5.875*cos((7*360/7)+90)},{5.875*sin((7*360/7+90))}) node {HORC};
    \draw ({5.875*cos((6*360/7)+90)},{5.875*sin((6*360/7+90))}) node {RDRC};
    \draw ({5.875*cos((5*360/7)+90)},{5.875*sin((5*360/7+90))}) node {HOSZ};
    \draw ({5.875*cos((4*360/7)+90)},{5.875*sin((4*360/7+90))}) node {NPOL};
    \draw ({5.875*cos((3*360/7)+90)},{5.875*sin((3*360/7+90))}) node {NDEF};
    \draw ({5.875*cos((2*360/7)+90)},{5.875*sin((2*360/7+90))}) node {RDSZ};
    \draw ({5.875*cos((1*360/7)+90)},{5.875*sin((1*360/7+90))}) node {RCSZ};


    \fill[pink] ({(20-11.1)/2*cos((7*360/7)+90)},{(20-11.1)/2*sin((7*360/7)+90)}) circle (.25);
    \fill[red] ({(20-12.7)/2*cos((7*360/7)+90)},{(20-12.7)/2*sin((7*360/7)+90)}) circle (.125);

    \fill[pink] ({(20-14.3)/2*cos((6*360/7)+90)},{(20-14.3)/2*sin((6*360/7)+90)}) circle (.25);
    \fill[red] ({(20-16)/2*cos((6*360/7)+90)},{(20-16)/2*sin((6*360/7)+90)}) circle (.125);

    \fill[pink] ({(20-11.2)/2*cos((5*360/7)+90)},{(20-11.2)/2*sin((5*360/7)+90)}) circle (.25);
    \fill[red] ({(20-14)/2*cos((5*360/7)+90)},{(20-14)/2*sin((5*360/7)+90)}) circle (.125);

    \fill[pink] ({(20-11.7)/2*cos((4*360/7)+90)},{(20-11.7)/2*sin((4*360/7)+90)}) circle (.25);
    \fill[red] ({(20-13)/2*cos((4*360/7)+90)},{(20-13)/2*sin((4*360/7)+90)}) circle (.125);

    \fill[pink] ({(20-18)/2*cos((3*360/7)+90)},{(20-18)/2*sin((3*360/7)+90)}) circle (.25);
    \fill[red] ({(20-15.7)/2*cos((3*360/7)+90)},{(20-15.7)/2*sin((3*360/7)+90)}) circle (.125);
    
    \fill[pink] ({(20-12)/2*cos((2*360/7)+90)},{(20-12)/2*sin((2*360/7)+90)}) circle (.25);
    \fill[red] ({(20-15)/2*cos((2*360/7)+90)},{(20-15)/2*sin((2*360/7)+90)}) circle (.125);

    \fill[pink] ({(20-12.7)/2*cos((1*360/7)+90)},{(20-12.7)/2*sin((1*360/7)+90)}) circle (.25);
    \fill[red] ({(20-14.5)/2*cos((1*360/7)+90)},{(20-14.5)/2*sin((1*360/7)+90)}) circle (.125);
    
    \fill[red,opacity=.5] ({(20-12.7)/2*cos((7*360/7)+90)},{(20-12.7)/2*sin((7*360/7)+90)}) -- ({(20-16)/2*cos((6*360/7)+90)},{(20-16)/2*sin((6*360/7)+90)}) -- ({(20-14)/2*cos((5*360/7)+90)},{(20-14)/2*sin((5*360/7)+90)}) -- ({(20-13)/2*cos((4*360/7)+90)},{(20-13)/2*sin((4*360/7)+90)}) -- ({(20-15.7)/2*cos((3*360/7)+90)},{(20-15.7)/2*sin((3*360/7)+90)}) -- ({(20-15)/2*cos((2*360/7)+90)},{(20-15)/2*sin((2*360/7)+90)}) -- ({(20-14.4)/2*cos((1*360/7)+90)},{(20-14.4)/2*sin((1*360/7)+90)});
    
    \end{tikzpicture}
      \begin{tikzpicture}[thin,>=stealth,scale=.625]
    
    \foreach \x in {1,2,3,4,5,6,7}
    \draw (0,0) -- ({5*cos((\x*360/7)+90)},{5*sin((\x*360/7+90))});
    
    \foreach \x in {1,2,3,4,5}
    \draw ({\x*cos((1*360/7)+90)},{\x*sin((1*360/7+90))}) -- ({\x*cos((2*360/7)+90)},{\x*sin((2*360/7+90))}) -- ({\x*cos((3*360/7)+90)},{\x*sin((3*360/7+90))}) -- ({\x*cos((4*360/7)+90)},{\x*sin((4*360/7+90))}) -- ({\x*cos((5*360/7)+90)},{\x*sin((5*360/7+90))}) -- ({\x*cos((6*360/7)+90)},{\x*sin((6*360/7+90))}) -- ({\x*cos((7*360/7)+90)},{\x*sin((7*360/7+90))}) -- ({\x*cos((1*360/7)+90)},{\x*sin((1*360/7+90))});
    
    \draw (0,{sin((4*360/7)+90))-.25}) node {$18$};
    \draw (0,{2*sin((4*360/7)+90))-.25}) node {$16$};
    \draw (0,{3*sin((4*360/7)+90))-.25}) node {$14$};
    \draw (0,{4*sin((4*360/7)+90))-.25}) node {$12$};
    \draw (0,{5*sin((4*360/7)+90))-.25}) node {$10$};
    
    \draw ({5.875*cos((7*360/7)+90)},{5.875*sin((7*360/7+90))}) node {HORC};
    \draw ({5.875*cos((6*360/7)+90)},{5.875*sin((6*360/7+90))}) node {RDRC};
    \draw ({5.875*cos((5*360/7)+90)},{5.875*sin((5*360/7+90))}) node {HOSZ};
    \draw ({5.875*cos((4*360/7)+90)},{5.875*sin((4*360/7+90))}) node {NPOL};
    \draw ({5.875*cos((3*360/7)+90)},{5.875*sin((3*360/7+90))}) node {NDEF};
    \draw ({5.875*cos((2*360/7)+90)},{5.875*sin((2*360/7+90))}) node {RDSZ};
    \draw ({5.875*cos((1*360/7)+90)},{5.875*sin((1*360/7+90))}) node {RCSZ};


    \fill[pink] ({(20-14.6)/2*cos((7*360/7)+90)},{(20-14.6)/2*sin((7*360/7)+90)}) circle (.25);
    \fill[red] ({(20-16.2)/2*cos((7*360/7)+90)},{(20-16.2)/2*sin((7*360/7)+90)}) circle (.125);

    \fill[pink] ({(20-16.3)/2*cos((6*360/7)+90)},{(20-16.3)/2*sin((6*360/7)+90)}) circle (.25);
    \fill[red] ({(20-17.2)/2*cos((6*360/7)+90)},{(20-17.2)/2*sin((6*360/7)+90)}) circle (.125);

    \fill[pink] ({(20-12.1)/2*cos((5*360/7)+90)},{(20-12.1)/2*sin((5*360/7)+90)}) circle (.25);
    \fill[red] ({(20-15)/2*cos((5*360/7)+90)},{(20-15)/2*sin((5*360/7)+90)}) circle (.125);

    \fill[pink] ({(20-12.3)/2*cos((4*360/7)+90)},{(20-12.3)/2*sin((4*360/7)+90)}) circle (.25);
    \fill[red] ({(20-13.6)/2*cos((4*360/7)+90)},{(20-13.6)/2*sin((4*360/7)+90)}) circle (.125);

    \fill[pink] ({(20-18)/2*cos((3*360/7)+90)},{(20-18)/2*sin((3*360/7)+90)}) circle (.25);
    \fill[red] ({(20-15.7)/2*cos((3*360/7)+90)},{(20-15.7)/2*sin((3*360/7)+90)}) circle (.125);
    
    \fill[pink] ({(20-13)/2*cos((2*360/7)+90)},{(20-13)/2*sin((2*360/7)+90)}) circle (.25);
    \fill[red] ({(20-16)/2*cos((2*360/7)+90)},{(20-16)/2*sin((2*360/7)+90)}) circle (.125);

    \fill[pink] ({(20-13.4)/2*cos((1*360/7)+90)},{(20-13.4)/2*sin((1*360/7)+90)}) circle (.25);
    \fill[red] ({(20-15.1)/2*cos((1*360/7)+90)},{(20-15.1)/2*sin((1*360/7)+90)}) circle (.125);
    
    \fill[red,opacity=.5] ({(20-16.2)/2*cos((7*360/7)+90)},{(20-16.2)/2*sin((7*360/7)+90)}) -- ({(20-17.2)/2*cos((6*360/7)+90)},{(20-17.2)/2*sin((6*360/7)+90)}) -- ({(20-15)/2*cos((5*360/7)+90)},{(20-15)/2*sin((5*360/7)+90)}) -- ({(20-13.6)/2*cos((4*360/7)+90)},{(20-13.6)/2*sin((4*360/7)+90)}) -- ({(20-15.7)/2*cos((3*360/7)+90)},{(20-15.7)/2*sin((3*360/7)+90)}) -- ({(20-16)/2*cos((2*360/7)+90)},{(20-16)/2*sin((2*360/7)+90)}) -- ({(20-15.1)/2*cos((1*360/7)+90)},{(20-15.1)/2*sin((1*360/7)+90)});
    
    \end{tikzpicture}
\end{center}
  \caption{Contributions to the $^{40-42}\text{Ca}^{19+}$ ground-state isotope shifts in the $g$ factor (left) and the dimensionless energy level $E$ (right). The large pink dots are the contributions, the small red dots are their uncertainties. The graph is in logarithmic scale, digits indicate negative powers of $10$.\\
  HORC: higher-order recoil. RDRC: radiative recoil. HOSZ: higher-order finite nuclear size. NPOL: nuclear polarization. NDEF: nuclear deformation. RDSZ: radiative finite nuclear size. RCSZ: finite nuclear size recoil. \label{fig:RadarRed}}
\end{figure*}
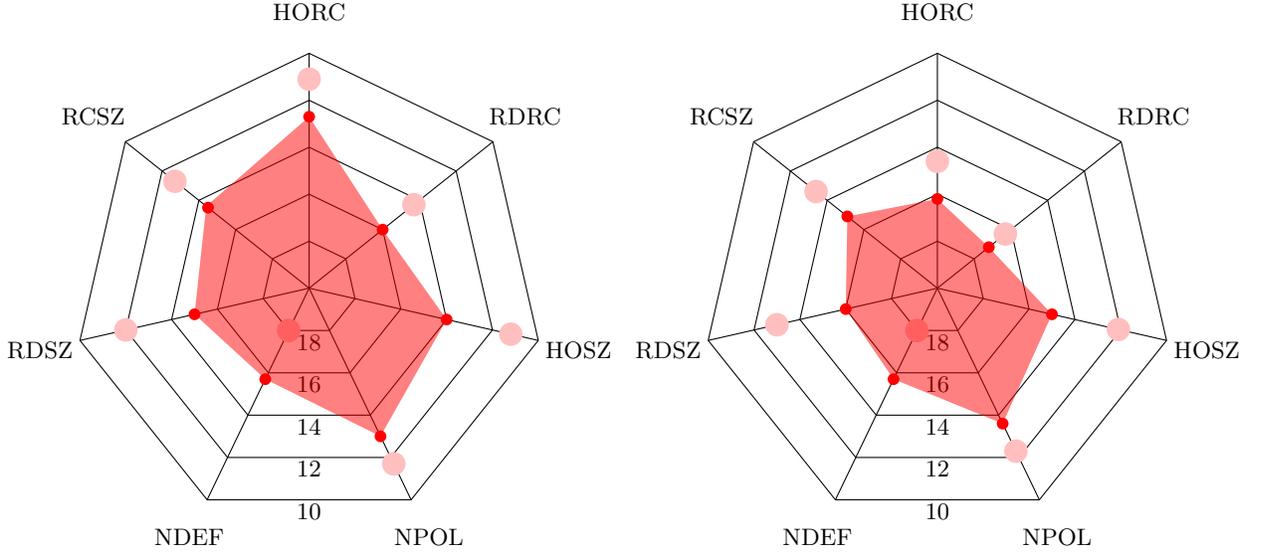
\begin{figure*}
\begin{center}
  \begin{tikzpicture}[thin,>=stealth,scale=.625]
    
    \foreach \x in {1,2,3,4,5,6,7}
    \draw (0,0) -- ({5*cos((\x*360/7)+90)},{5*sin((\x*360/7+90))});
    
    \foreach \x in {1,2,3,4,5}
    \draw ({\x*cos((1*360/7)+90)},{\x*sin((1*360/7+90))}) -- ({\x*cos((2*360/7)+90)},{\x*sin((2*360/7+90))}) -- ({\x*cos((3*360/7)+90)},{\x*sin((3*360/7+90))}) -- ({\x*cos((4*360/7)+90)},{\x*sin((4*360/7+90))}) -- ({\x*cos((5*360/7)+90)},{\x*sin((5*360/7+90))}) -- ({\x*cos((6*360/7)+90)},{\x*sin((6*360/7+90))}) -- ({\x*cos((7*360/7)+90)},{\x*sin((7*360/7+90))}) -- ({\x*cos((1*360/7)+90)},{\x*sin((1*360/7+90))});
    
    \draw (0,{sin((4*360/7)+90))-.25}) node {$26$};
    \draw (0,{2*sin((4*360/7)+90))-.25}) node {$24$};
    \draw (0,{3*sin((4*360/7)+90))-.25}) node {$22$};
    \draw (0,{4*sin((4*360/7)+90))-.25}) node {$20$};
    \draw (0,{5*sin((4*360/7)+90))-.25}) node {$18$};
    
    \draw ({5.875*cos((7*360/7)+90)},{5.875*sin((7*360/7+90))}) node {HORC};
    \draw ({5.875*cos((6*360/7)+90)},{5.875*sin((6*360/7+90))}) node {RDRC};
    \draw ({5.875*cos((5*360/7)+90)},{5.875*sin((5*360/7+90))}) node {HOSZ};
    \draw ({5.875*cos((4*360/7)+90)},{5.875*sin((4*360/7+90))}) node {NPOL};
    \draw ({5.875*cos((3*360/7)+90)},{5.875*sin((3*360/7+90))}) node {NDEF};
    \draw ({5.875*cos((2*360/7)+90)},{5.875*sin((2*360/7+90))}) node {RDSZ};
    \draw ({5.875*cos((1*360/7)+90)},{5.875*sin((1*360/7+90))}) node {RCSZ};


    \fill[cyan] ({(26-17.0)/2*cos((7*360/7)+90)},{(26-17.0)/2*sin((7*360/7)+90)}) circle (.25);

    \fill[cyan] ({(26-20.2)/2*cos((6*360/7)+90)},{(26-20.2)/2*sin((6*360/7)+90)}) circle (.25);

    \fill[cyan] ({(26-18.1)/2*cos((5*360/7)+90)},{(26-18.1)/2*sin((5*360/7)+90)}) circle (.25);

    \fill[cyan] ({(26-18.5)/2*cos((4*360/7)+90)},{(26-18.5)/2*sin((4*360/7)+90)}) circle (.25);

    \fill[cyan] ({(26-24)/2*cos((3*360/7)+90)},{(26-24)/2*sin((3*360/7)+90)}) circle (.25);
    
    \fill[cyan] ({(26-18.3)/2*cos((2*360/7)+90)},{(26-18.3)/2*sin((2*360/7)+90)}) circle (.25);

    \fill[cyan] ({(26-18.5)/2*cos((1*360/7)+90)},{(26-18.5)/2*sin((1*360/7)+90)}) circle (.25);
    
    
    \end{tikzpicture}
      \begin{tikzpicture}[thin,>=stealth,scale=.625]
    
    \foreach \x in {1,2,3,4,5,6,7}
    \draw (0,0) -- ({5*cos((\x*360/7)+90)},{5*sin((\x*360/7+90))});
    
    \foreach \x in {1,2,3,4,5}
    \draw ({\x*cos((1*360/7)+90)},{\x*sin((1*360/7+90))}) -- ({\x*cos((2*360/7)+90)},{\x*sin((2*360/7+90))}) -- ({\x*cos((3*360/7)+90)},{\x*sin((3*360/7+90))}) -- ({\x*cos((4*360/7)+90)},{\x*sin((4*360/7+90))}) -- ({\x*cos((5*360/7)+90)},{\x*sin((5*360/7+90))}) -- ({\x*cos((6*360/7)+90)},{\x*sin((6*360/7+90))}) -- ({\x*cos((7*360/7)+90)},{\x*sin((7*360/7+90))}) -- ({\x*cos((1*360/7)+90)},{\x*sin((1*360/7+90))});
    
    \draw (0,{sin((4*360/7)+90))-.25}) node {$12$};
    \draw (0,{2*sin((4*360/7)+90))-.25}) node {$11$};
    \draw (0,{3*sin((4*360/7)+90))-.25}) node {$10$};
    \draw (0,{4*sin((4*360/7)+90))-.25}) node {$9$};
    \draw (0,{5*sin((4*360/7)+90))-.25}) node {$8$};
    
    \draw ({5.875*cos((7*360/7)+90)},{5.875*sin((7*360/7+90))}) node {HORC};
    \draw ({5.875*cos((6*360/7)+90)},{5.875*sin((6*360/7+90))}) node {RDRC};
    \draw ({5.875*cos((5*360/7)+90)},{5.875*sin((5*360/7+90))}) node {HOSZ};
    \draw ({5.875*cos((4*360/7)+90)},{5.875*sin((4*360/7+90))}) node {NPOL};
    \draw ({5.875*cos((3*360/7)+90)},{5.875*sin((3*360/7+90))}) node {NDEF};
    \draw ({5.875*cos((2*360/7)+90)},{5.875*sin((2*360/7+90))}) node {RDSZ};
    \draw ({5.875*cos((1*360/7)+90)},{5.875*sin((1*360/7+90))}) node {RCSZ};


    \fill[blue] ({(13-8.3)*cos((7*360/7)+90)},{(13-8.3)*sin((7*360/7)+90)}) circle (.125);

    \fill[blue] ({(13-11.5)*cos((6*360/7)+90)},{(13-11.5)*sin((6*360/7)+90)}) circle (.125);

    \fill[blue] ({(13-10.5)*cos((5*360/7)+90)},{(13-10.5)*sin((5*360/7)+90)}) circle (.125);

    \fill[blue] ({(13-9.5)*cos((4*360/7)+90)},{(13-9.5)*sin((4*360/7)+90)}) circle (.125);

    \fill[blue] ({(13-11.7)*cos((3*360/7)+90)},{(13-11.7)*sin((3*360/7)+90)}) circle (.125);
    
    \fill[blue] ({(13-10.3)*cos((2*360/7)+90)},{(13-10.3)*sin((2*360/7)+90)}) circle (.125);

    \fill[blue] ({(13-11.3)*cos((1*360/7)+90)},{(13-11.3)*sin((1*360/7)+90)}) circle (.125);
    
    \fill[blue,opacity=.5] ({(13-8.3)*cos((7*360/7)+90)},{(13-8.3)*sin((7*360/7)+90)}) -- ({(13-11.5)*cos((6*360/7)+90)},{(13-11.5)*sin((6*360/7)+90)}) -- ({(13-10.5)*cos((5*360/7)+90)},{(13-10.5)*sin((5*360/7)+90)}) -- ({(13-9.5)*cos((4*360/7)+90)},{(13-9.5)*sin((4*360/7)+90)}) -- ({(13-11.7)*cos((3*360/7)+90)},{(13-11.7)*sin((3*360/7)+90)}) -- ({(13-10.3)*cos((2*360/7)+90)},{(13-10.3)*sin((2*360/7)+90)}) -- ({(13-11.3)*cos((1*360/7)+90)},{(13-11.3)*sin((1*360/7)+90)});
    
    \end{tikzpicture}
\end{center}
  \caption{Contributions to the SM King nonplanarity (left) and to the bound on the NP coupling constant $4\pi\alpha_{\mathrm{HP}}$ (right) for ground-state $^{40-42-44-48}\text{Ca}^{19+}$ in the vanishing boson mass limit. The large cyan dots are the contributions, the small blue dots are due to the uncertainties of the subleading SM nuclear corrections. The graph is in logarithmic scale, digits indicate negative powers of $10$.\\
  HORC: higher-order recoil. RDRC: radiative recoil. HOSZ: higher-order finite nuclear size. NPOL: nuclear polarization. NDEF: nuclear deformation. RDSZ: radiative finite nuclear size. RCSZ: finite nuclear size recoil. \label{fig:RadarBlue}}
\end{figure*}
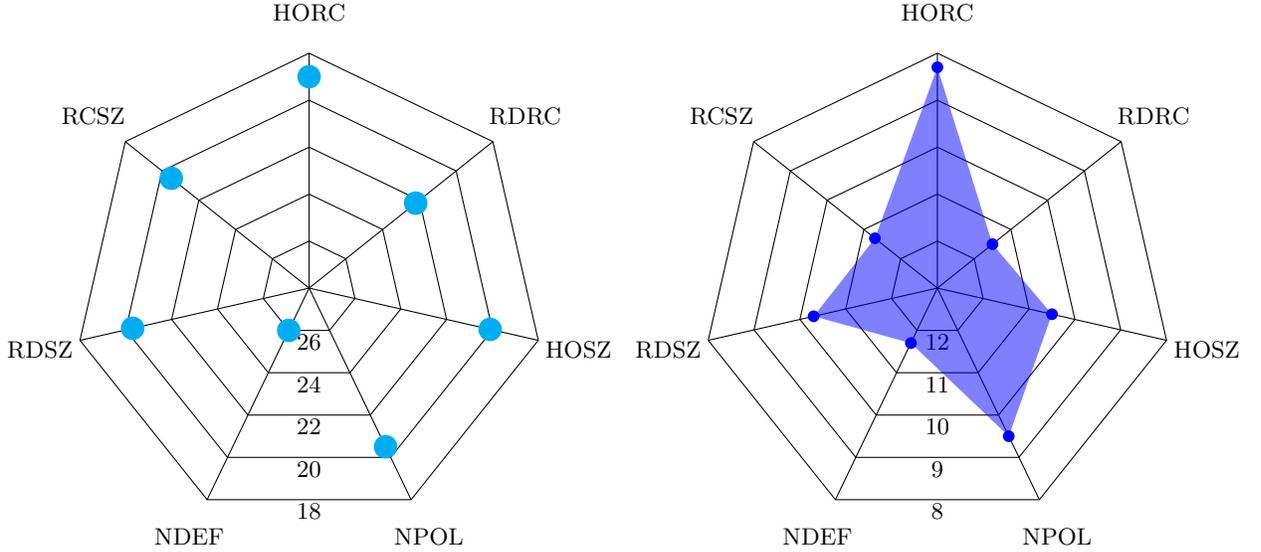
 For that reason, we introduce the square-mass King representation. Let us rewrite the isotope shift~(\ref{eq:FullNP}) as
 \begin{equation} \label{eq:SquareMass}
  g_i^{AA'}=K_i\,\mu_{AA'}+K_i^{\left(2\right)}\,\beta_{AA'}+H_i\,\delta R_{AA'}^{2\gamma_i}+s_{i\,AA'}+n_{i\,AA'},
\end{equation}
with $\beta_{AA'}=1/M_A^2-1/M_{A'}^2$. Here, we have written the quadratic recoil contribution explicitly. We now introduce the square-mass normalised quantities $G_{i\beta}^{AA'}\equiv g_i^{AA'}/\beta_{AA'}$. We also introduce $\delta_{\beta}^{AA'}\equiv\delta R_{AA'}^{2\gamma}/\beta_{AA'}$, $S_{i\beta\,AA'}\equiv s_{i\beta\,AA'}/\beta_{AA'}$ and $N_{i\beta\,AA'}\equiv n_{i\beta\,AA'}/\beta_{AA'}$, as well as $\mu_{\beta}^{AA'}\equiv\mu_{AA'}/\beta_{AA'}$ and $h_{\beta}^{AA'}\equiv\left(A-A'\right)/\beta_{AA'}$. Now, with vectors constructed as in Eq.~(\ref{eq:Vectors}), we can proceed to similar manipulations, to obtain the square-mass King nonplanarity
\begin{equation} \label{eq:NonBeta}
  \mathcal{N}_\beta\equiv\frac{1}{2}\left(\mathbf{G}_{1\beta}\times\mathbf{G}_{2\beta}\right)\cdot\mathbf{A}_{\mathrm{tt}},
\end{equation}
\begin{widetext}
\begin{multline} \label{eq:ArghNArrowsBeta}
  \mathcal{N}_\beta=-\frac{\mathbf{A}_{\mathrm{tt}}}{2}\cdot\left[\mathbf{G}_{1\beta}\times\left(\frac{H_2}{H_1}\mathbf{S}_{1\beta}-\mathbf{S}_{2\beta}\right)-\left(1\leftrightarrow2\right)\right]+\frac{\alpha_{\mathrm{HP}}}{2}\left(\mathbf{A}_{\mathrm{tt}}\times\mathbf{h}_{\beta}\right)\cdot\left[\left(\frac{H_2}{H_1}X_1-X_2\right)\mathbf{G}_{1\beta}-\left(1\leftrightarrow2\right)\right]\\
  -\frac{\mathbf{A}_{\mathrm{tt}}}{2}\cdot\left[\mathbf{G}_{1\beta}\times\left(\frac{H_2}{H_1}K_{1\beta}-K_{2\beta}\right)\bm{\mu}_\beta-\left(1\leftrightarrow2\right)\right].
\end{multline}
Of course, this expression is not equal to that given in Eq.~(\ref{eq:ArghNArrows}), but if one compares the terms one by one, it is seen that the first two summands on the r.h.s. of Eq.~(\ref{eq:ArghNArrowsBeta}) have an equivalent in Eq.~(\ref{eq:ArghNArrows}), while the third one does not. That third term is the contribution to $\mathcal{N}_\beta$ from the leading-order mass shift. In the absence of NP, and if one neglects the subleading nuclear corrections, this third summand survives as a contribution to the square-mass King nonplanarity (\ref{eq:NonBeta}), which should hence not be expected to vanish when NP is absent and the subleading nuclear corrections are neglected. We can then use Eq.~(\ref{eq:ArghNArrowsBeta}) to solve for $\alpha_{\mathrm{HP}}$, and set bounds on the NP coupling constant through
\begin{equation} \label{eq:PropErrorAlphaBeta}
  \sigma_{\alpha_{\mathrm{HP}}}=\sqrt{\sum_{i,j}\left(\frac{\partial\alpha_{\mathrm{HP}}}{\partial G_{i\beta\left(j\right)}}\right)^2\left(\Delta G_{i\beta\left(j\right)}\right)^2}+\sqrt{\sum_{i,j}\left(\frac{\partial\alpha_{\mathrm{HP}}}{\partial S_{i\beta\left(j\right)}}\right)^2\left(\Delta S_{i\beta\left(j\right)}\right)^2}+\sqrt{\sum_{i,j}\left(\frac{\partial\alpha_{\mathrm{HP}}}{\partial K_{i}}\right)^2\left(\Delta K_{i}\right)^2}.
\end{equation}
\end{widetext}
The uncertainty on the NP coupling constant, due to the leading nuclear recoil correction, is reduced by four orders of magnitude compared to that generated by the higher-order nuclear recoil correction. The contribution from the six other subleading nuclear corrections remain essentially unchanged from those shown in Fig.~\ref{fig:RadarBlue}, so that this new square-mass King representation is a clear improvement.

\section{Direct tests} \label{sec:Direct}

Despite all the refinements introduced in Sec.~\ref{sec:Indirect}, searching for NP through King planarity analysis turns out not to be the most efficient approach for few-electron ions. The King approach is needed for many-electron systems, in particular because for these systems, the electron-interaction contributions to the nuclear recoil, the so-called specific mass shift, cannot be computed accurately. The King approach, in which the nuclear recoil is eliminated, is thus useful for such systems. With few-electron ions, on the other hand, calculations can be carried out more comprehensively and more accurately, which indicates the interest of a direct approach to the search for NP.

\begin{figure}[b]
\begin{center}
  \begin{tikzpicture}[very thick,>=stealth]
    \draw[->] (0,-1.5) -- (0,1.5);
    \draw (-.25,1.25) node {$g$};
    \draw (.75,-.75) -- (1.25,-.75);
    \draw (1.625,-.75) node {\large $g_{\mathrm{XP}}$};
    \draw (2.25,.25) -- (2.75,.25);
    \draw (3.125,.25) node {\large $g_{\mathrm{TH}}$};
    \draw[|-|] (1,-1.125) -- (1,-.3875);
    \draw[|-|] (2.5,-.5) -- (2.5,1);
    \draw[dashed] (0,-1.125) -- (5,-1.125);
    \draw[dashed] (0,1) -- (5,1);
    \draw[<-] (4,-1.125) -- (4,-3*.1125);
    \draw[->] (4,3*.1) -- (4,1);
    \draw (4,-.006125) node {\large $g_{\mathrm{HP}}^{\left(\mathrm{max}\right)}$};
\end{tikzpicture}
\end{center}
  \caption{Schematic representation of the direct method of setting bounds on New Physics. The experimental $g_{\mathrm{XP}}$ and theoretical $g_{\mathrm{TH}}$ values of the $g$ factor are given with their uncertainties. The largest discrepancy allowed by the error bars gives the largest possible contribution $g_{\mathrm{HP}}^{\left(\mathrm{max}\right)}$ to the $g$ factor from one of the New Physics candidates (Higgs portal). This directly allows the setting of bounds on the New Physics parameters.  \label{fig:ErrorBars}}
\end{figure}
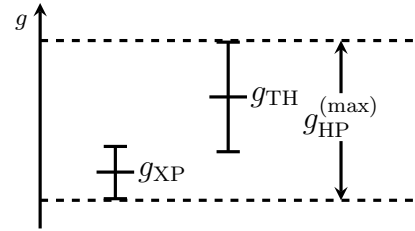

In this approach, we use the small discrepancy between theory and experiment to derive bounds on hypothetical New Physics (NP) contributions. In some cases, theory and experiment may agree within their respective error bars, while, in other cases, the error bars may not overlap. However, which case  is irrelevant to our specific concern here, since the maximum discrepancy between theory and experiment allowed by the error bars may be set as the maximum hypothetical NP contribution in either case (see Fig.~\ref{fig:ErrorBars}). This approach was explored further in Ref.~\cite{FifthForceG} in the case of the $g$ factor of few-electron ions, but, implemented on a single $g$ factor, it necessitates challenging improvements in the calculation of QED corrections, and also interelectronic-interaction contributions, to spectroscopic quantities. Many such corrections are under active consideration~\cite{CzarneckiLetter,CzarneckiLog,SikoraTwoLoop,CakirLiB,TwoLoopMLoop}.

A better approach consists of the same direct comparison between experiment and theory, but implemented on a single isotope shift~\cite{Vincent_single_arxiv}. Thus, radiative contributions from QED largely drop out, reducing the amount of necessary theory input. 
As is clear from the foregoing discussion, the bound that can be set on NP is limited by the uncertainty $\delta s_{AA'}$ on the subleading nuclear corrections. In the Higgs portal scenario, the NP contribution to the isotope shift of the $g$ factor or energy level is $n_{AA'}=\alpha_{\mathrm{HP}}\left(A-A'\right)X$. The difference in the number of neutrons will typically be of the order of unity, and as can be seen from Eqs.~(\ref{eq:ENP}) and (\ref{eq:GNP}), the electronic coefficient is roughly $Z\alpha$ in the light boson limit, so that for $Z>10$, a given uncertainty on the subleading nuclear corrections will yield a 10 times larger contribution, very roughly speaking, to the bound on $\alpha_{\mathrm{HP}}$. This is to be contrasted with the corresponding contribution in the King approach, which, as can be
seen in Figs.~\ref{fig:RadarRed} and \ref{fig:RadarBlue}, is several orders of magnitude larger than the corresponding uncertainties on the subleading nuclear corrections.

\section{Conclusion} \label{sec:Ccl}

In this work, the search for New Physics with few-electron ions through King linearity analysis has been pushed to what is arguably its farthest extent. We have modified the King formalism to explicitly account for subleading Standard Model nuclear corrections, which contribute to isotope shifts and to King nonlinearity. This yielded a King linearity violation test which can be carried out in the very-high-precision experimental regime. 
Indeed, this test enables detection of New Physics contributions to spectroscopic quantities which would be smaller than the subleading nuclear corrections. 
We also introduced the square-mass King representation to successfully suppress uncertainties from the higher-order recoil. 
Even with all these refinements, the bounds which can be set with this approach remain less competitive than bounds obtained with the spectroscopy of many-electron ions~\cite{FifthForce,NonLinXPYb,LinXPCa} putting forward a simpler approach with a single isotope pair and no King analysis can yield more stringent bounds with the same amount of theory input~\cite{Vincent_single_arxiv}. 
Finally, our results stress the importance of a detailed analysis of the subleading corrections to the King's plot, and would hopefully stimulate for same investigation for the other systems \cite{PossibleForces,FifthForce,FlambaumIso,NonLinXPYb,LinXPCa}  being used for the search of the New Physics beyond the Standard Model.

\bibliography{Biblio}
\end{document}